\documentclass{article}

\usepackage[english]{babel}

\usepackage[letterpaper,top=2cm,bottom=2cm,left=3cm,right=3cm,marginparwidth=1.75cm]{geometry}

\usepackage{amsmath}
\usepackage{graphicx}
\usepackage[colorlinks=true, allcolors=blue]{hyperref}

\title{What is the Metaverse? An Immersive Cyberspace and Open Challenges} 
\author{Lik-Hang Lee\footnote{corresponding author: lhleeac@connect.ust.hk; L.H. Lee is affiliated with KAIST, S. Korea; P. Zhou is affiliated with USTC, China; T. Braud and P. Hui are affiliated with HKUST, China.}, Pengyuan Zhou, Tristan Braud, Pan Hui}

\begin{document}
\maketitle

\begin{abstract}
The Metaverse refers to a virtual-physical blended space in which multiple users can concurrently interact with a unified computer-generated environment and other users, which can be regarded as the next significant milestone of the current cyberspace. This article primarily discusses the development and challenges of the Metaverse. 
We first briefly describe the development of cyberspace and the necessity of technology enablers. Accordingly, our bottom-up approach highlights three critical technology enablers for the Metaverse: networks, systems, and users. Also, we highlight a number of indispensable issues, under technological and ecosystem perspectives, that build and sustain the Metaverse.  
\end{abstract}

\section{The Metaverse: the Next Immersive 3D Internet} 
The Metaverse was first coined by \textit{Neal Stephenson}'s science-fiction novel entitled `\textbf{Snow Crash}' in 1992. The novel depicts the main characters with their avatars living in a virtual-physical blended world, characterized by omnipresent virtual entities on top of our physical surroundings. In such a blended world, people can perform various immersive tasks. Prominent examples include gathering with their friends remotely, working with their workmates collaboratively, and co-experiencing virtual events (e.g., concerts and shopping). 
In other words, diversified digital or virtual contents from cyberspace will go beyond 2D screens in our existing cyberspace and gradually penetrate into 3D environments. 

Coincidentally, the imagined world, as mentioned above, aligns with \textit{Mark Weiser}'s vision of ubiquitous computing in 1991: computing services will be embedded into multitudinous aspects of our life, and users can access the virtual content anytime and anywhere. 
With such a compelling vision, over the last three decades, the landscape of ubiquitous computing has been advanced by burgeoning computing devices, including laptop computers, smartphones, the Internet of Things, and smart wearables.
Meanwhile, the current cyberspace has evolved drastically, and the latest attempts proceed to offer services and digital experiences to human users through virtual environments, such as Augmented Reality (AR) and Virtual Reality (VR), according to Milgram and Kishino's Reality-Virtuality Continuum~\cite{Milgram1994ATO}. Albeit nobody can tell exactly what the Metaverse will deliver when it comes true, the recent pre-metaverse applications do likely pinpoint AR and VR on smartphones that serve as the primary testbed of immersive user experiences. For instance, Pokémon Go has become the most popular AR application on ubiquitous devices, remarkably with 1 billion download, while Google Cardboard brings VR multimedia (e.g., YouTube VR) to mass audiences. 

\section{A Quick Walkthrough of the Metaverse}

\begin{figure}
\centering
\includegraphics[width=\textwidth]{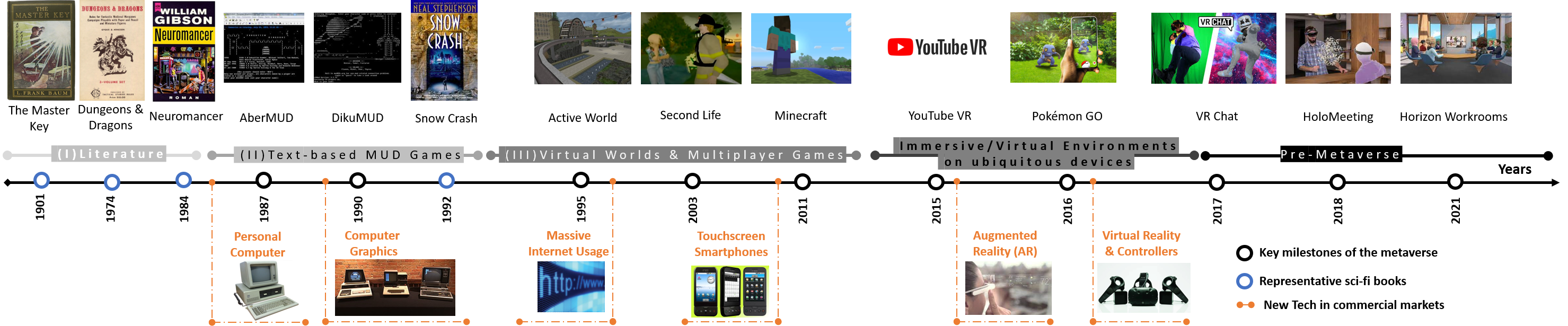}
\caption{The centennial chronicles of the Metaverse (1901 -- 2021).}
\label{fig:review}
\end{figure}

Obviously, the technology always serves as the essential building block that catalyzes the next milestone of cyberspace in the metaverse era, as shown in Figure~\ref{fig:review}.
Before we reach the quo status (i.e., the pre-Metaverse (V)), we have experienced four consecutive stages, namely (I) Literature, (II) Text-based virtual worlds, (III) 3D Virtual Worlds, and (VI) AR and VR on Ubiquitous devices.
In the first stage, literature becomes the endless space of imagination for technologies. The most representative novels include `\textit{The Master Key}', `\textit{Dungeons and Dragons}', `\textit{Neuromancer}', and `\textit{Snow Crash}', which project the immersive experiences even though the technologies were not yet ready. 
Since the advent of personal computers, Multi-User Dungeon (MUD) games were proliferating in 1985 (II) that support text-based user interaction. Next, once the computer graphics were available on personal computers, the very first 3D virtual world, namely \textit{Active World}, was launched in 1995 (III). The massive population of the Internet has further made multiplayer 3D virtual worlds, such as \textit{Second Life} and \textit{Minecraft}. Moreover, the rise of smartphones in the late 2000s (VI) became the foundation of mobile AR and VR applications mentioned above. After 2015 (V), the commercialization of AR and VR headsets, e.g., Google Glass and Meta Oculus, enables users enclosed by computer-mediated environments, as well as sees digital overlays in virtual-physical blended environments, respectively, e.g., VR Chat, Horizon Workrooms, and HoloMeeting. 

It is worth noting that the pandemic since 2020 can be regarded as one of the most extensive ``experiments" in history -- do people accept the further movement of various functions of life into virtual environments? Obviously, yes. Certain first-mover technology giants, e.g., Meta and Microsoft, promote virtual collaborative environments for everyone with affordable AR and VR ubiquitous devices, including end consumers and enterprises.
Therefore, we foresee that the next progressions of cyberspace featured by immersiveness, i.e., immersive cyberspace, have to consider other technology enablers. 
The remaining paragraphs in the article serve as a reality check of technology enablers and their challenges. We follow a bottom-up approach to discuss three critical aspects, inherited the viewpoints of ubiquitous computing in particular: Network and Mobile Edge, System Interoperability, and User-centric Immersive Design, representing network, system, to human users. We acknowledge that the scope of technology enablers for the Metaverse (e.g., Artificial Intelligence, Blockchain, Computer Vision, Robotics~\cite{Lee2021AllON}) can extend farther than our discussion. 

\section{Rising Demands for Edge}
In the Metaverse, it is essential to guarantee an immersive feeling for the user to provide the same level of experience as reality. One of the most critical factors is latency, e.g., motion to photon (MTP) latency\footnote{MTP latency is the amount of time between the user's action and its corresponding effect to be reflected on the display screen.}. Researchers have found that MTP latency needs to be below the human perceptible limit to allow users to interact with holographic augmentations seamlessly and directly~\cite{10.1145/1012551.1012559}. For instance, in the registration process of AR, large latency often results in virtual objects lagging behind the intended position, which may cause sickness and dizziness. As such, reducing latency is critical for the Metaverse, especially in scenarios where real-time data processing is demanded, e.g., real-time AR interaction with the physical world, such as AR surgeries.
On the other hand, the Metaverse often requires too intensive computation for mobile devices and greatly limits the applicability. 

\subsection{Offloading}
To this end, offloading is often used to relieve the computation and memory burden at the cost of additional networking latency. Therefore, a balanced tradeoff is crucial to make the offloading process transparent to the users, which unfortunately is not easy. For example, rendering a locally navigable viewport larger than the headset's field of view is necessary to balance out the networking latency during offloading. However, there is a tension between the required viewport size and the networking latency: longer latency requires a larger viewport and streaming more content, resulting in even longer latency. Therefore, a solution with physical deployment improvement may be more realistic than pure resource orchestration.

Due to the variable and unpredictable high latency, cloud offloading cannot always reach the optimal balance and causes long-tail latency performance, which impacts user experience~\cite{40801}. Thus a complementary solution is demanded to guarantee a seamless and immersive user experience in the Metaverse.
Edge computing, which computes, stores, and transmits the data physically closer to end-users and their devices, can reduce the user-experienced latency compared with cloud offloading. As early as 2009, Satyanarayanan~\emph{et al.}~\cite{cloudlet} recognized that deploying powerful cloud-like infrastructure just one wireless hop away from mobile devices, i.e., the so-called cloudlet, could change the game. 
The latest mobile AR frameworks have employed edge-based solutions 
to improve the performance of Metaverse applications. For instance, EdgeXAR offers a mobile AR framework taking the benefits of edge offloading to provide lightweight tracking with 6 Degree of Freedom and hides the offloading latency from the user's perception~\cite{edgexar}. EAVVE proposes a novel cooperative AR vehicular perception system facilitated by edge servers to reduce the overall offloading latency and makes up for the insufficient in-vehicle computational power~\cite{9163287}. 

\subsection{Scale to Outdoor}
By offloading to an edge server (e.g., a high-end PC), users can enjoy a more interactive and immersive experience at higher framerates without sacrificing the detail of immersive environments. However, such solutions are constrained to indoor environments with limited user mobility.
To allow truly and fully omnipresent metaverse experience, seamless outdoor mobility is critical. With the development of 5G and 6G, Multi-access edge computing (MEC) is expected to boost metaverse user experience by providing standard and universal edge offloading services one-hop away from the cellular-connected user devices, e.g., AR glasses. Not only can it reduce the round-trip-time (RTT) of packet delivery, but also it opens a door for near real-time orchestration for multi-user interactions. MEC is crucial for outdoor metaverse services to comprehend the detailed local context and orchestrate intimate collaborations among nearby users or devices. For instance, 5G MEC servers can manage nearby users' AR content with only one-hop packet transmission and enable real-time user interaction for social AR applications such as `Pokémon Go'. 
Employing MEC to improve metaverse experience has acquired academic attention and has been deployed by some metaverse companies to improve user experience\footnote{\url{https://nianticlabs.com/blog/niantic-planet-scale-ar-5g-urban-legends/}}.

\section{Interoperability, Openness, and Universality}

At the time of writing this article, the metaverse projects and proposals primarily consist of individual projects under the total control of a single entity. Although such fragmentation allows for experimenting with concepts and ideas in an exploratory era, such a model is not sustainable long-term. We even argue that interoperability and openness will be the two primary make-or-break components in the worldwide adoption of the Metaverse. 

\subsection{Considering the Web as an Example}

In many aspects, the Metaverse can be considered the next evolution from the Web as a new form of content creation and sharing. From the very beginning, the Web was designed for decentralization and interoperability~\cite{berners2015history}. Any person or entity with sufficient technical skills, a computer, and access to the Internet can deploy a website. At its core, the Web does not rely on any central entity to operate, and more centralized services (e.g., Domain Name Service) are only convenient additions to a completely decentralized technical foundation. As a matter of fact, the first website ever deployed was running on Tim Berners-Lee's personal machine at the CERN. This low barrier to entry, combined with the user keeping complete control over what is published, gave users ownership of the medium, and led to the success of the Web. 

Interoperability and openness were instrumental in making the Web the ubiquitous technology it has become nowadays. However, it also led to multiple actors developing their own versions of the HTML language and rendering engines. In 1994, Tim Berners-Lee founded the World Wide Web Consortium (W3C)\footnote{World Wide Web Consortium - \url{https://www.w3.org/}} in an attempt to encourage compatibility across the industry, and, more importantly, the vendor-neutrality of the Web. Nowadays, as a standardization organism, the W3C drafts and establishes the standards for the World Wide Web, under the governance of its members.

\subsection{Achieving Interoperability in the Metaverse}

The Metaverse should build on a model similar to the Web, where a multitude of virtual worlds hosted by various entities can be accessed by the users through devices and browsers made by diverse companies and organizations. However, the Metaverse presents characteristics that significantly affect this operation. 

The Web was a technology entirely created and managed by a single entity before being released to the wider community. Comparatively, there have been many attempts to create the Metaverse since the late 1990s, starting with the French ``Deuxième Monde"\footnote{Mémoires du Deuxième Monde - \url{https://www.bimondiens.com/}} (1997) and the more famous "Second Life"\footnote{Second Life - \url{https://secondlife.com/}} (2003). Contrary to the Web, most of these attempts were designed under the model of massively multiplayer online games, where a single company keeps absolute control over the content published on their platform. More recent blockchain-based proposals such as Decentraland\footnote{Welcome to Decentraland - \url{https://decentraland.org/}} go one step further, where content is stored on distributed servers (IPFS), and ownership is tracked through a blockchain. However, the web platform through which the user accesses the metaverse world remains dependent on its parent entity, and there is little interoperability with other platforms. 
Besides, we argue that IPFS takes away the possibility for users to delete data, going against the "right to be forgotten" raised by several bodies around the world. With the proliferation of non-interoperable metaverse projects, it is necessary now more than ever to establish standardization organisms that can address the following aspects:

\begin{itemize}
    \item The possibility for anybody to deploy a server and host a virtual world connected to the rest of the Metaverse.
    \item Accessing the Metaverse through any device and browser that respects an established specification (client-based rendering).
    \item Keeping track of digital asset ownership across multiple servers and clients.
    \item Enabling a single avatar to interact with other avatars across servers.
    \item Allowing users to create, display, trade, and delete digital assets in the Metaverse.
\end{itemize}

These characteristics are critical for users to take ownership of the Metaverse and drive ubiquitous adoption, paving the way to a ubiquitous metaverse.

\subsection{The Case of Immersive Technologies}

Immersive technologies such as augmented reality and virtual reality (AR and VR) are at the core of both ubiquitous computing and the Metaverse. In ubiquitous computing, they allow for better contextualization when visualizing data, while in the Metaverse, they reinforce the user's immersion in the virtual world. These technologies are often considered as a spectrum ranging from reality to complete virtuality, with applications blending digital content with a physical context to various degrees. With the Metaverse being in its infancy, there is no strict definition of how much these technologies should interface with our physical world. This uncertainty will resolutely affect interoperability, as some metaverse applications may be strongly anchored within a physical location~\cite{braud2022scaling}, while others may be entirely virtual~\cite{duan2021metaverse}, accessible through a traditional desktop computer. A study on developing an AR metaverse campus~\cite{braud2022scaling} suggests replicating the physical environment into a fully virtual experience through digital twins, allowing for geolocalised experiences to be accessible to remote users. However, similar to the issues mentioned above, only a standardization organism can clearly define how to establish such duality.

\section{Towards Metaverse-User Interaction}

\begin{figure}
\centering
\includegraphics[width=\textwidth]{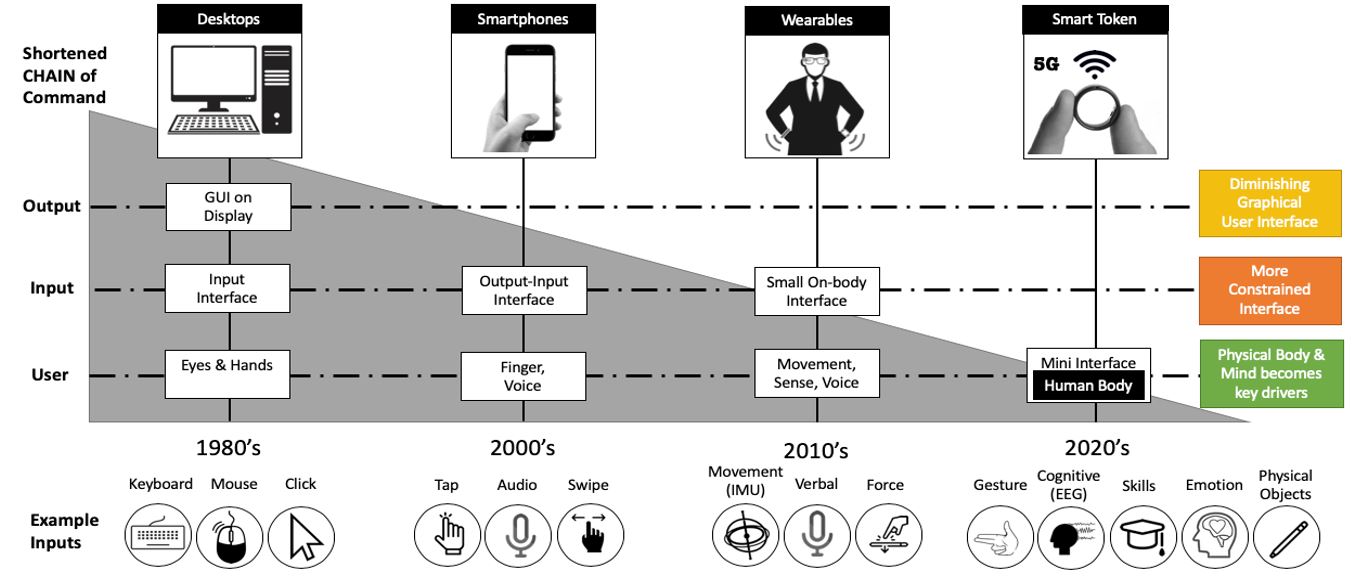}
\caption{The trend of mobile interfaces is projected to become centric of the user's body and smaller in size, perhaps eventually invisible.}
\label{fig:trend}
\end{figure}

Considering the Metaverse will eventually penetrate into every aspect of daily routines, users should own channels to convert their intentions into actions in immersive environments anytime and anywhere~\cite{Lee2020UbiPointTN}. The existing AR and VR applications on smartphones shed light on the wide adoption of virtual content overlaid on the physical world. The new generation of metaverse devices, e.g., AR and VR headsets, offer better immersion than smartphones. Nonetheless, nowadays, headsets lack efficient user input methods. Thus, user interaction with virtual contents in the Metaverse becomes difficult and tedious, under the premise that interfaces are getting smaller in size (Figure~\ref{fig:trend}). Numerous research prototypes aim to improve novel input channels featured with user mobility. For instance, Google's Jacquard\footnote{\url{https://atap.google.com/jacquard/}} leverages embedded sensors into garments, which augments our daily clothes as a large touch interface for touchscreen-like user interaction. Another type of mobile input is wearable addendums, including wristbands, gloves, and finger-worn devices, which extend users' bodies as the center of user interaction with virtual contents. Although emerging numbers of mobile input solutions have been proposed in recent years, there still exists a noticeable gap in input capability between the sedentary solution like the keyboard and mouse duos and the aforementioned mobile input solutions~\cite{Lee2020TowardsAR}. 

Moreover, the information in physical worlds can be considered countless information, while augmentation inside the considerably small headsets' field-of-views is challenging~\cite{seen-to-unseen}. As such, we have to optimize the presentation of virtual content. Otherwise, metaverse users with a naive approach to virtual content management will encounter information overflow, thus significantly extra time in selecting augmentation. One prominent strategy is context awareness, e.g., the users, environments, and social dynamics. The edge AIs, e.g., recommendation systems, can parse user context and give the most relevant augmentation~\cite{A2W-LAM2021}.
In addition, the current research on user contexts primarily focuses on the users' five senses that are straightforward and factual. Instead, additional efforts should be made to enable the user-centric Metaverse to understand the abstract yet hard-to-quantify feelings on top of the five senses, for instance, a house with a dark and purple background vs. a daunted house (abstract feeling). A high granularity of context-awareness can offer more precise and personalized services in the Metaverse.

Metaverse users will leave many user interaction traces in such virtual-physical blended cyberspace, although edge computing can improve not only user experiences, as mentioned, and user privacy. 
Apart from the edge infrastructure, user privacy and design space remain unexplored. 
As it is expected that users will perform various virtual activities, and hence any inappropriate augmentation can lead to new privacy threats. For example, users may leave their `augmented' dialogues in the virtual public space. Visually augmenting every line of dialogue with multiple users can pose privacy threats~\cite{kumar2021theophany}. Information flows, perhaps driven by contextual integrity, should be reinforced by virtual space owners. Multiple factors contribute to the user and data contexts, such as the recipient, locations, data sensitivity, and so on. However, we know very little about the design of information flows for highly diversified augmentation, not to mention the unknown forms of augmentation to appear in the Metaverse. 

\section{Outlook and Challenges}
The Metaverse emphasizes a new height of user experiences across the physical and virtual worlds, and therefore requires concurrent efforts from networking, system, and user-centric aspects. As a final note, we highlight several immediate issues of three technological aspects that would become the bottleneck of implementing the `ubiquitous' Metaverse at scale. 

The first challenge is the delay-transparent offloading for metaverse applications. To compensate for the limited mobile device capacity, real-time offloading is necessary for delay-sensitive metaverse applications. Cloud has been a dominating offloading choice for years, but its delay has been found to fail to meet the aforementioned MTP requirement. Privacy issue rising with the untrustworthiness of deep learning widely used in the cloud is also a major concern. The fast-developing 5G and 6G network technologies are showing promising potential for an extremely low last-mile delay to enable delay-transparent edge offloading. Hence the Metaverse can see a future of ubiquitous service provision that allows users freely enjoy the immersive experience in the wild. 

Second is the societal and technological obstacles toward an interoperable metaverse. As the technologies that will support the Metaverse are not yet well-defined, it is currently difficult to see a standard emerging. Similar to the Web, the Metaverse should allow any entity to create and host their own virtual world, that users can visit with their avatars. Digital content creation and ownership across virtual worlds will be the primary technological challenge to enable such intercompatibility. On the societal aspects, we expect to witness a similar trend to open-source software, with communities pushing for interoperability for either humanistic (free access to the medium) or pragmatic (openness drives innovation). At the same time, other entities will strive to establish monopolies over the Metaverse.

The final aspect addresses the users' throughput rates and user perception of virtual environments. There is a vast gap in throughput rates between mobile input devices and the traditional yet sedentary interfaces. Also, content management and display on AR/VR headsets, as well as a better understanding of users' abstract feelings, are still underexplored. 


Similar to the development of the Internet, the Metaverse, also known as the \textit{immersive Internet}, will take decades to build. More importantly, once the virtual environments are widely deployed in our surroundings, the Metaverse will take a considerable long period to iterate and refine its ecosystems. Therefore, we have to consider various ecosystem issues that can impact the sustainability of the Metaverse in the long run~\cite{Lee2021AllON}. The potential issues include avatar behaviors in the wild, trust in multi-avatar identity, content creations, cross-generational contents, cancel culture, decentralized design of virtual economy and governance, digital humanity, user diversity and fairness, online addiction, user data ownership and ethics, user safety with AR and VR, stewardship, and accountability.

\bibliographystyle{abbrv}
\bibliography{main}

\end{document}